# New orbital angular momentum multiplexing strategy: beyond the capacity limit of free-space optical communication

Wenxiang Yan,[1,2] Yuan Gao,[1,2] Xian Long,[1,2] Zheng Yuan,[1,2] Zhi-Cheng Ren,[1,2] Xi-Lin Wang,[1,2] Jianping Ding,[1,2,3,*] and Hui-Tian Wang[1,2,*]

[1]National Laboratory of Solid State Microstructures and School of Physics, Nanjing University, Nanjing 210093, China

[2]Collaborative Innovation Center of Advanced Microstructures, Nanjing University, Nanjing 210093, China

[3]Collaborative Innovation Center of Solid-State Lighting and Energy-Saving Electronics, Nanjing University, Nanjing 210093, China

*Corresponding author: jpding@nju.edu.cn; htwang@nju.edu.cn

**Abstract:** Free-space optical (FSO) communication can exploit mode-division multiplexing using orthogonal spatial modes of Laguerre-Gaussian beams, such as orbital angular momentum (OAM) modes, wherein OAM multiplexing offers potentially infinite information capacity due to the arbitrary quantization of OAM. Combined with polarization-division multiplexing and wavelength-division multiplexing, OAM multiplexing is a promising solution for future capacity demands. However, the practically addressable number of spatial subchannels is severely limited by the receiver size and the rapid beam expansion with increasing mode order and communication distance. Based on the intrinsic and distinctive property that the divergent degree of the innermost ring of a Laguerre-Gaussian beam is significantly slower than that of the beam cross-section during propagation, here we propose theoretically and demonstrate experimentally a novel communication strategy—innermost-ring-dominated OAM (IRD-OAM) multiplexing—that can overcome these limits and achieve up to 1238% capacity of conventional OAM multiplexing in a canonical FSO link system without any additional hardware modifications. Alternatively, our strategy can also enable longer communication distance (403% of that for conventional OAM multiplexing), or smaller receiver (26.9% in size compared to conventional OAM multiplexing), while maintaining the same capacity as conventional OAM multiplexing. Our work will hasten the development of future FSO communications with ultra-high capacity, ultra-long distance and highly-integrated devices for deep-space, near-Earth and Earth-surface applications.

## Introduction

Increasing capacity for harnessing and processing information has been a longstanding goal for scientists and engineers[1]. Multiplexing of optical degrees of freedom, such as polarization and wavelength, has been widely used to enhance radiofrequency and optical communication capacity for decades[2-4]. Mode-division multiplexing offers a new possibility to multiply capacity by using orthogonal spatial modes as independent information channels[5-10], such as conventional orbital angular momentum (OAM) modes with zero radial index ($p = 0$), which are a subset of Laguerre-Gaussian (LG) beams. For example, a free-space optical (FSO) link that combines mode-division multiplexing with $Q$ orthogonal modes, polarization-division multiplexing with two orthogonal polarization states, and wavelength-division multiplexing with $T$ wavelengths, along with being encoded with 100 Gbit/s quadrature phase-shift keying data, can achieve an aggregate capacity of $Q \times 2 \times T \times 100$ Gbit/s (ref. [11]), which can even reach up to several Pbit/s (ref. [8]) and potentially enable future deep-space and near-Earth optical communications with ultra-high capacity and spectral efficiency[12-17]. A major challenge for FSO links with mode-division multiplexing is that the beam divergence due to diffraction and finite receiver size limit the number of addressable subchannels $Q$ and the link distance[18-21] (Fig. 1a). Theoretically an OAM multiplexing provides infinite information capacity because of arbitrary quantization of OAM, even though it only exploits the azimuthal degree of freedom ($\varphi$) of spatial resource in an FSO communication system. Furthermore, the mode-division multiplexing using LG beams with the azimuthal and radial indices ($m$, $p$), called LG beam multiplexing, exploits both two spatial degrees of freedom ($r$, $\varphi$) and thus can make better use of always limited spatial resources, providing hope in breaking through the limits of communication capacity that are constrained by current FSO communication systems[22].

Here, we discovered and demonstrated that the innermost ring of the LG beam possesses superior dynamic transmission characteristics: a significantly smaller quality factor ($M^2$), self-healing capability with the aid of sidelobes, and better identification ability owing to larger intensity and phase gradient than the sidelobes. Therefore, we propose to utilize the innermost ring of the LG beam for information transfer and recovery, named innermost-ring-dominated OAM (IRD-OAM) multiplexing technique. Our results showed that the IRD-OAM multiplexing is very appealing for high-capacity communication via FSO systems with a limited-size receiver. In contrast to the OAM multiplexing and LG beam multiplexing, which only make trade-offs in the "triangle of frustration" (capacity, link distance, receiver miniaturization), the IRD-OAM multiplexing can go significantly beyond the original communication limit imposed by the FSO system by integrating multidimensional physical resources. Using the canonical FSO link system, the IRD-OAM multiplexing can achieve 1238% of the capacity limit for the OAM multiplexing; additionally, given the same capacity as the OAM multiplexing, the use of the IRD-OAM multiplexing enables a further extension of communication distance (403% of that for the OAM multiplexing) and allows a much smaller receiver size (26.9% of that for the OAM multiplexing). This new IRD-OAM multiplexing strategy showed its potential for the future data communication with ultra-high capacity, ultra-long distance and highly-integrated devices.

## Result

### The superiority of the innermost ring

Figure 1a illustrates the schematic view of FSO communication using data-carrying LG beams. The complex amplitude distribution of a normalized LG beam carrying OAM of $m\hbar$ per photon in cylindrical coordinate system ($r, \varphi, z$) is represented by

$$\mathrm{LG}_{m,p}(r,\varphi,z) = \sqrt{\frac{2p!}{\pi(|m|+p)!}} \frac{1}{w(z)} \left[\frac{\sqrt{2}r}{w(z)}\right]^{|m|} L_p^{|m|}\left[\frac{2r^2}{w^2(z)}\right] \exp(-\frac{r^2}{w^2(z)}) \qquad (1\text{-}1)$$
$$\times \exp\left[ikz + ik\frac{r^2}{2R(z)} + im\varphi - i(|m|+2p+1)\zeta(z)\right],$$

where $L_p^m$ denotes the Laguerre polynomial with the azimuthal index $m$ (or topological charge) and the radial index $p$, $k=2\pi/\lambda$ is the wavenumber with $\lambda$ being the wavelength, and

$$w(z) = w_0\sqrt{1+\left(\frac{z}{z_0}\right)^2}, \ R(z) = z\left[1+\left(\frac{z_0}{z}\right)^2\right], \ \zeta(z) = \tan^{-1}\left(\frac{z}{z_0}\right), \ z_0 = \frac{\pi w_0^2}{\lambda}, \ w_0 = \sqrt{\frac{\lambda z_0}{\pi}}, \qquad (1\text{-}2)$$

with $w_0$ denoting the beam waist. The root-mean-squared waist radius (defines the size of the LG beam) $r_{LG}(z) = \sqrt{|m|+2p+1}\,w(z)$ and divergence angle $\theta_{LG} = \lim_{z\to\infty} \mathrm{d}r_{LG}(z)/\mathrm{d}z = \sqrt{|m|+2p+1}\,\theta_0$ describe the

free-space propagation property of the LG beam and are vital for the FSO communication[23], where $\theta_0$ is the fundamental Gaussian mode divergence. The beam quality factor ($M^2$) (ref. [24]), defined as the ratio between the space–bandwidth products of the LG beam, $r_{LG}(0)\theta_{LG}$, and of the fundamental Gaussian mode, $w_0\theta_0$, characterizes the propagation dynamics based on the inherent uncertainty principle between the beam size and divergence[25]

$$M_{LG}^2(m,p) = \frac{r_{LG}(0)\theta_{LG}}{w_0\theta_0} = |m| + 2p + 1. \quad (2)$$

Because $r_{LG}(z) = M_{LG}(m,p)w(z)$ and $\theta_{LG} = M_{LG}(m,p)\theta_0$, different LG beams with the same $M^2_{LG}$ always hold the same size upon propagating, i.e., retaining the same propagation dynamics. In addition, in the most OAM multiplexing systems, OAM modes can be rendered by $LG_{m,0}(r, \varphi, z)$, i.e., a subset of LG beams with $p = 0$, whose beam size $r_{OAM}(z) = \sqrt{|m|+1}w(z)$ and quality factor $M^2_{OAM}(m) = |m|+1$ are both the smallest for each mode with OAM of $m\hbar$ per photon.

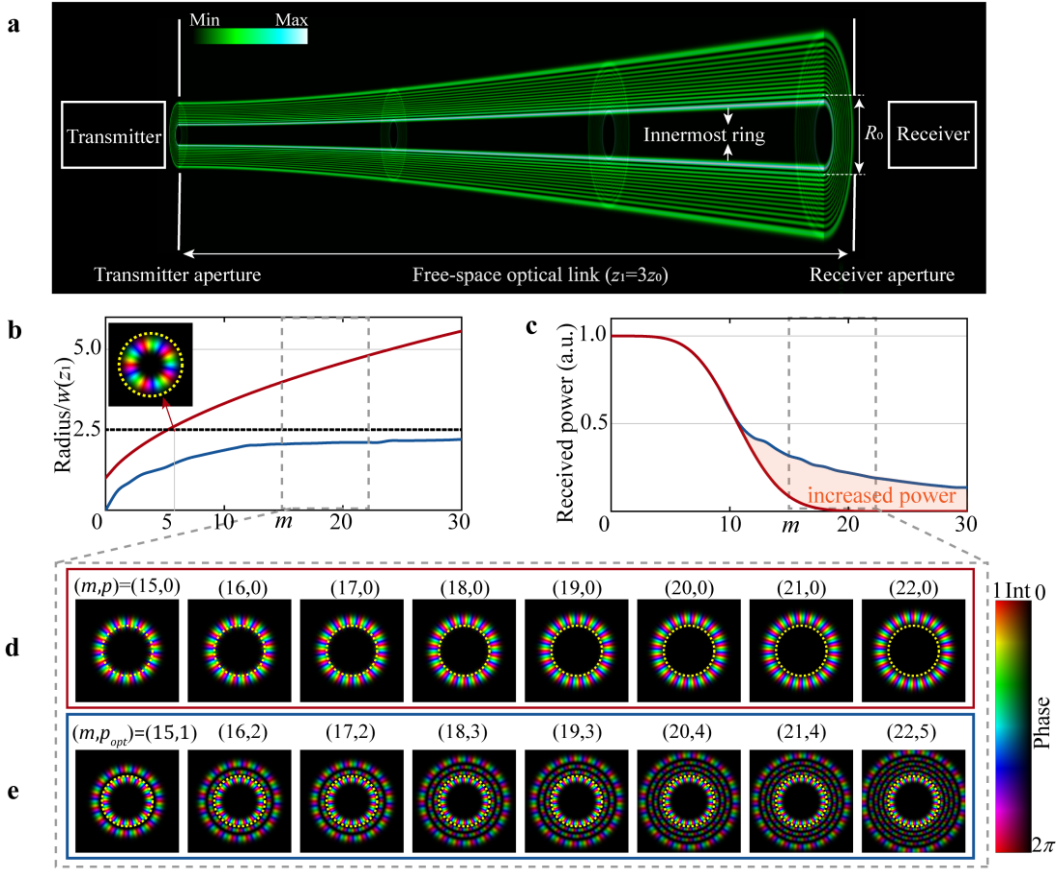

**Fig. 1. Challenge of FSO communication using data-carrying LG beams. a** A limited-size receiver of radius $R_0 = 2.5\,w(z_1)$ at a link distance $z_1(=3z_0)$ obstructs the LG beam due to rapid divergence, but allows the innermost ring to pass through. **b** Dependence of the required receiver aperture radius on the azimuthal index $m$: the root-mean-squared waist radii $r_{OAM}(z_1)$ of $LG_{m,0}(r, \varphi, z_1)$ (red curve), the innermost ring radii of $LG_{m,p_{opt}(m)}(r,\varphi,z_1)$ (blue curve), and the limited-size aperture of radius $R_0 = 2.5w(z_1)$ (black-dashed line). The subscript "opt" of $p$ in **d** denotes the optimal radial index $p$ for a given $m$ for brevity. The inset in **b** shows the complex distribution of $LG_{5,0}(r, \varphi, z_1)$ and the yellow-dotted circle represents the receiver aperture. **c** Received power of $LG_{m,0}(r, \varphi, z_1)$ (red curve) and of $LG_{m,p_{opt}(m)}(r,\varphi,z_1)$ (blue curve), with the orange shadow area indicating the increasing power between the two curves. **d** Complex distributions of $LG_{m,0}(r, \varphi, z_1)$ and **e** of $LG_{m,p_{opt}(m)}(r,\varphi,z_1)$ with $m$ ranging from 15 to 22.

A key design challenge for the mode-division-multiplexing-based FSO links is the divergence of LG beams due to the diffraction as they propagate, as illustrated in Fig. 1**a**, because the receiver window always has a limited size, which limits the number ($Q$) of independent subchannels[18-21]. As an illustration, we consider the FSO link system in Fig. 1 with the receiver aperture of the radius $R_0 = 2.5w(z_1)$ at $z_1$ and the transmitter at $z = 0$. As is well known that the LG modes with $r_{LG}(z_1) \leq R_0$ can be orthogonally demultiplexed in terms of $(m, p)$ and thus are regarded as independently addressable communication subchannels. Obviously, as shown in Fig. 1**b-d** and represented by the red curves, the OAM modes $LG_{m,0}$ have $r_{LG}(z_1) > R_0$ for $m > 5$ and thus cannot afford the subchannels of this FSO system.

We aim to determine the conditions under which FSO systems can transmit data-carrying LG beams. To achieve this, a framework needs to be established for calculating $Q$ factor of different physical FSO link systems. Analogous to the commonly used beam quality factor $M^2$ for characterizing the beam propagation dynamics, we introduce "the system quality factor" $S^2$ as a means to describe the transmission capability of a canonical FSO link system with limited-size apertures[22]. We define $S^2$ as the ratio between the space–bandwidth products of the FSO link system ($2R_0 \times 2NA/\lambda$) and the fundamental Gaussian system ($4/\pi$), where $NA$ is the numerical aperture of circular apertures of the receiver. Due to the inherent relationship between size and divergence, such a choice of system quality factor takes advantage of the aperture size and link distance, similar to how the beam quality factor is defined by the available space and bandwidth. The system quality factor $S^2$ characterizes those FSO link systems whose receiver aperture radius at $z_1$ are $R_0 = Sw(z_1)$, namely, the radius of receiver aperture should "diverge" when increasing the link distance $z_1$, akin to the propagation dynamics of LG beams. When the beam quality factor $M^2 \leq S^2$, i.e. meeting $r_{LG}(z_1) = M_{LG}w(z_1) \leq R_0 = Sw(z_1)$, this beam can pass through this system without clipping. As a result, the number of solutions for $M^2_{LG}(m,p) \leq S^2$ determines $Q$ of an FSO link system that uses the LG beam multiplexing as the information carrier, which is given by $Q_{LG} \approx 0.5\text{floor}[S^2](\text{floor}[S^2]+1)$, where floor[·] is the floor function that rounds down to obtain an integer value. Likewise, the $Q$ factor of the OAM multiplexing which satisfes $M^2_{OAM}(m) \leq S^2$ is $Q_{OAM} \approx 2\text{floor}[S^2]+1$ (ref. [22]). For instance, in the system illustrated in Fig. 1 whereby $R_0 = 2.5w(z_1)$, the value of $S^2$ is 6.25, thereby yielding $Q_{LG}$ and $Q_{OAM}$ of 21 and 13, respectively.

In practice, the receiver aperture radius is fixed at $R_0$, resulting in $S^2 = [R_0/w(z_1)]^2$. Thus, the system quality factor decreases with an increase in link distance $z_1$ or a decrease in receiver aperture $R_0$, ultimately leading to a reduction in the transmitting capability of those diverging LG beams, thereby severely limiting independently addressable subchannels $Q$ and information capacity. However, we have found that the innermost ring of LG beam exhibits superior dynamic transmission characteristics owing to its significantly smaller quality factor (as well as size and divergence) and self-healing capability with sidelobes. As a result, information recovery from the innermost ring is feasible even with a limited-size receiver.

After analyzing Eq. (1), we have obtained the following analytical solution for the innermost vortex ring radius of the LG beam (i.e., the radius of the brightest ring, instead of the root-mean-squared waist radius) (see Supplementary text 1 for details)

$$r_{IR}(z) = \frac{C_1\sqrt{(|m|+1)|m|} + C_2/2}{\sqrt{|m|+2p+1}} w(z), \qquad (3)$$

where $C_1 = 0.5207$ and $C_2 = 0.7730$ are two constant coefficients. The subscript "$IR$" denotes the parameters related to the innermost ring for brevity. The divergence angle is given by

$$\theta_{IR} = \lim_{z \to \infty} dr_{IR}(z)/dz = \frac{C_1\sqrt{(|m|+1)|m|} + C_2/2}{\sqrt{|m|+2p+1}} \theta_0. \qquad (4)$$

Consequently, the quality factor defined by the innermost ring of the LG beam can be written as

$$M_{IR}^2(m,p) = \frac{r_{IR}(0)\theta_{IR}}{w_0\theta_0} = \frac{\left(C_1\sqrt{(|m|+1)|m|} + C_2/2\right)^2}{|m|+2p+1}. \tag{5}$$

As $r_{IR}(z) = M_{IR}(m,p)w(z)$ and $\theta_{IR} = M_{IR}(m,p)\theta_0$, the sizes of innermost rings with the same $M^2_{IR}$ remain constant (i.e., maintain the same transmission dynamics) during propagation. Such rings can be regarded as a kind of "perfect" vortices that have OAM-independent radii and weakest dependence on propagation (i.e., long "depth of focus"). In contrast, in the case of conventional perfect vortices, their radii are OAM-independent only in the vicinity of a plane (focal plane) while cannot preserve to be OAM-independent and quickly diverge when propagating[26,27].

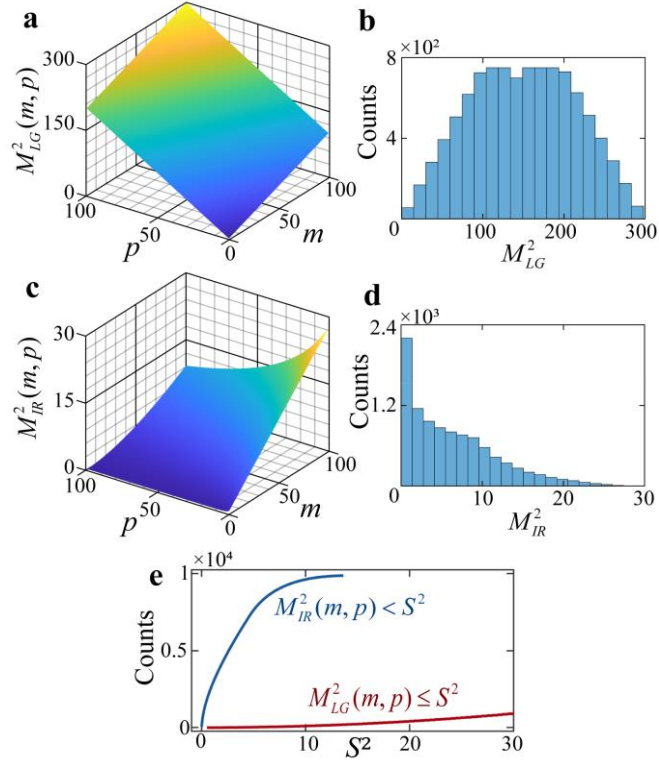

**Fig. 2. Quality factors of LG beams, innermost rings and FSO link systems.** Quality factors of **a** LG beams $M^2_{LG}(m,p)$ and **c** innermost rings $M^2_{IR}(m,p)$ for 10000 lowest orders ($m$ and $p$ equal 0, 1, …, 99 respectively; results for $m < 0$ are the same as those for $m > 0$ and are omitted here). The corresponding distribution histograms are shown in **b** and **d**. **e** Counts of solutions for $M^2_{LG}(m,p) \leq S^2$ (red curve) and $M^2_{IR}(m,p) < S^2$ (blue curve) for 10000 lowest orders.

According to Eqs. (3) and (4), $M^2_{IR}(m, p)$, $r_{IR}(z)$, and $\theta_{IR}$ increase with $|m|$ but decrease with $p$, whereas $M^2_{LG}(m, p)$, $r_{LG}(z)$, and $\theta_{LG}$ increase with both $|m|$ and $p$. Therefore, $M^2_{IR}(m, p)$, $r_{IR}(z)$ and $\theta_{IR}$ are much smaller than $M^2_{LG}(m, p)$, $r_{LG}(z)$, and $\theta_{LG}$ (including the subset $M^2_{OAM}(m)$, $r_{OAM}(z)$, and $\theta_{OAM}$) with the same $|m|$. We give a comparison between $M^2_{LG}(m, p)$ and $M^2_{IR}(m, p)$ in the 10000 lowest orders, as shown in Fig. 2**a-d**, showing that $M^2_{LG}(m, p)$ is dispersed within a range from 0 to 300, while $M^2_{IR}(m, p)$ is primarily distributed between 0 and 20. As a result, the propagation properties of the innermost ring, including the size, divergence, and quality factor, are far superior to those of the LG beam. For a practical system with definite quality factor $S^2$, the number of LG beams that can pass through the innermost ring (namely satisfying $M^2_{IR}(m, p) < S^2$) is several orders of magnitude greater than the number of LG beams that are required to pass entirely (fulfilling $M^2_{LG}(m,p) \leq S^2$), as illustrated in Fig. 2**e**. Hence, a new strategy for exceeding the communication capacity limit, without requiring additional hardware modifications to FSO link systems with a limited-size receiver, is feasible: retrieving information from the innermost ring, instead of recovering data from the entire beam. Furthermore, the innermost ring offers enhanced performances in both separating multiplexed beams with minimal modal crosstalk (via the drastic phase information) and recovering data with a high signal-to-noise ratio (SNR) (via the high intensity information), thus eliminating the design trade-off in FSO links[18]. Moreover, the innermost ring of the LG beam with $p \neq 0$ (as shown in Fig. 1**e**) has an intriguing and useful self-healing property: it tends to reconstitute its shape with the aid of sidelobes even after being severely impaired. The outer sidelobes act as a "reservoir" to store energy for the self-healing of the innermost ring after impairment, as shown in Supplementary text 2. This self-healing capability renders the innermost ring more resilient and sturdy against disturbance than OAM modes (i.e., LG beams with $p = 0$) and other regions of the LG beam. This capability further permits the adaption of this new strategy in optical communication in turbulent environments such as seawater and atmosphere.

Due to the rapid divergence with the mode order and communication distance, a limited-size receiver leads to a severe power loss in the mode-division multiplexing, such as OAM multiplexing and LG beam multiplexing. Specifically, for $m > 5$, LG beams (including OAM modes with $p = 0$) cannot pass through the receiver aperture. As the mode order approaches $m = 15$ they lose nearly all power, as depicted by the red curves in Figs. 1**b** and 1**c**. Nonetheless, our strategy of retrieving information from the innermost ring, which always passes through the receiver aperture with optimal radial index $p_{opt}(m)$ (as illustrated by the blue curve in Fig. 1**b**), allows us to maximize the total received power (as shown by the blue curves and orange shadow area in Fig. 1**c**).

## Exceeding the capacity limit by innermost-ring-dominated OAM multiplexing

Unlike the OAM multiplexing which uses only the azimuthal degree of freedom ($\varphi$), the LG beam multiplexing exploits both two spatial degrees of freedom ($r$, $\varphi$) of the system, thanks to the orthogonality of modes with $m$ and $p$ in Eq. (1). The limited-size receiver aperture truncates the $r$ dimension, which leads to $p$ non-orthogonal, while only $m$ remains to be orthogonal. Namely, only the orthogonality of OAM holds, as detailed in Supplementary text 3. Accordingly, $Q$ is determined by the number of $m$ solutions for $M^2_{IR}(m,p) < S$, thereby enabling the independently addressable and low-crosstalk subchannels in OAM multiplexing (see Supplementary texts 3 and 4 for details). To maximize the received power for each subchannel of the limited-size receiver, the radial index $p$ should be optimized for a given $m$. We define the receiving efficiency, $\eta(m, p)$, as the ratio of the power in the receiver aperture to that of the entire beam. Figure 3**a** shows the calculated efficiency versus ($m$, $p$) in an FSO link system with a given $S^2 = 6.25$ shown in Fig. 1. The dashed line in Fig. 3**a** represents the highest efficiency $\eta_{max}(m)$ for a given $m$ and the corresponding optimized $p_{opt}$ (in Fig. 3**b**). Figure 3**a** illustrates how to select an optimized radial index $p_{opt}(m)$ for a given $m$ that can significantly improve the receiving efficiency. Unlike conventional OAM multiplexing that uses modes ($m, p = 0$) to minimize the entire size of the beam, our new OAM multiplexing strategy achieves the highest receiving efficiency and minimizes power loss by utilizing the innermost rings of beams in modes ($m, p_{opt}$) for data recovery, named innermost-ring-dominated OAM (IRD-OAM) multiplexing. As shown in Fig. 3**c**, only the IRD-OAM multiplexing, rather than the OAM multiplexing or LG beam multiplexing, can be accomplished by an FSO link system that has a transmission capability of $S^2 = 6.25$, as we will explain in the following.

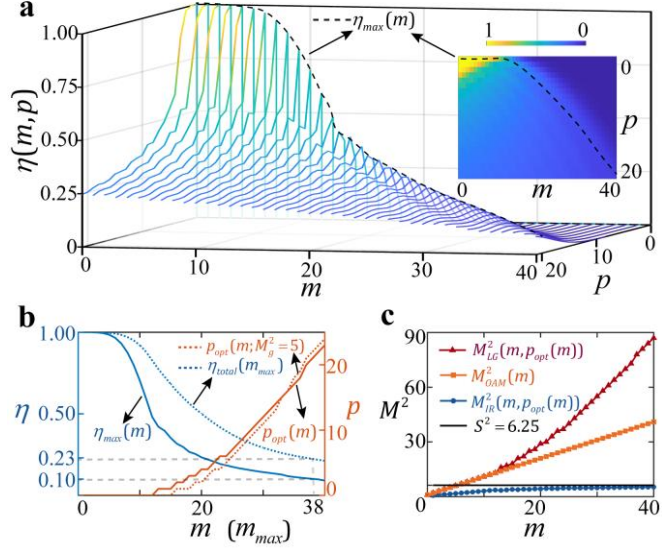

**Fig. 3. Reception efficiency of LG beam with $S^2$=6.25. a** Receiving efficiency $\eta(m, p)$ for $m$ in [0,40] and $p$ in [0, 23]; the dashed lines represent the highest efficiency $\eta_{max}(m)$ versus $(m, p)$. **b** Curves of $\eta_{max}(m)$ (blue), the total receiving efficiencies $\eta_{total}(m_{max})$ (blue-dotted), $p_{opt}(m)$ (orange), and $p_{opt}(m; M^2_g = 5)$ in Eq. (6) (orange-dotted); the gray dashed lines depict $\eta_{max}(38) = 0.10$, $\eta_{total}(38) = 0.23$. **c** $M^2_{LG}(m, p_{opt}(m))$, $M^2_{OAM}(m)$, and $M^2_{IR}(m, p_{opt}(m))$ (represented by red-triangle, orange-square, and blue-circle curves, respectively) versus the transmission capability of the FSO link system with $S^2$=6.25 (black line).

For low-order OAM modes with their beam size smaller than the receiver aperture (i.e. $M^2_{OAM}(m) \leq S^2$), we can select $p_{opt}(m) = 0$. For higher-order modes with larger beam sizes, we can increase $p$ to a critical value to decrease $r_{IR}$ in Eq. (3) such that the innermost ring of the beam can just fall within the receiver aperture, thereby reaching the maximum efficiency. This critical value is $p = p_{opt}(m)$, as seen in the modes $(m, p_{opt}(m))$ ($m$ ranging from 15 to 22) in Fig. 1**e**. The innermost rings truncated by the receiver aperture (the yellow-dotted circles in Fig. 1**e**) can be regarded as divergence-free "perfect" vortices, which have OAM-independent radius $r_{IR}(z)$ during propagation. In Fig. 3**c**, we compared the receiving ability of LG beam multiplexing, OAM multiplexing, and IRD-OAM multiplexing in an FSO link system with $S^2 = 6.25$, using the curves of $M^2_{LG}(m, p_{opt}(m))$, $M^2_{OAM}(m)$, and $M^2_{IR}(m, p_{opt}(m))$. We found that only $M^2_{IR}(m, p_{opt}(m))$ is always less than $S^2$, indicating that the innermost rings of different subchannels $(m, p_{opt}(m))$ can always pass through the receiver aperture. Nonetheless, the entire beams with $m > 5$ are highly confined by the aperture ($M^2_{LG} > M^2_{OAM} > S^2 = 6.25$ when $m > 5$), which is also manifested in Fig. 1**b**. To simplify the process of selecting $p_{opt}(m)$ for each independently addressable subchannel of the IRD-OAM multiplexing, we propose a straightforward method: select a global $M^2_g$ for all subchannels, and solve $p_{opt}(m; M^2_g)$ from Eq. (5), which is given by

$$p_{opt}(m; M^2_g) = \text{round}\left( \frac{\left(2C_1\sqrt{(|m|+1)|m|}+C_2\right)^2}{2M^2_g} - \frac{|m|+1}{2} \right), \tag{6}$$

where round(·) denotes rounding to the nearest integer. To avoid possible negative value of radial index $p$, the calculated $p_{opt}(m; M^2_g)$ from Eq. (6) for low-order OAM modes with $M^2_{IR}(m, 0) < M^2_g$ is set to be 0. The global factor $M^2_g$ should be smaller than $S^2$, so we chose $M^2_g = 5$ for the FSO link system with $S^2 = 6.25$. The resultant $p_{opt}(m; M^2_g = 5)$ was calculated and depicted in Fig. 3**b**. Note that the difference between $p_{opt}(m; M^2_g = 5)$ and $p_{opt}(m)$ is negligible, as shown in Fig. 3**b**. As such, the optimal radial index $p_{opt}(m)$ for a given $m$, obtained by searching for the highest efficiency $\eta_{max}(m)$ (see Fig. 3**a**), can be replaced in practice with the analytical expression of $p_{opt}(m; M^2_g)$ in Eq. (6).

The SNR is a fundamental metric for evaluating the performance of an optical communication system. In this context, the initial SNR [which is denoted by $N_0 = 10\lg(P_s/P_n)$] represents the ratio of the signal power ($P_s$) to the detector's noise power ($P_n$). In mode-division multiplexing like OAM multiplexing (Fig. 1**c**) and LG beam multiplexing (Fig. 3**a**), owing to the rapid divergence with the mode order and communication distance, a limited-size receiver causes severe loss of received power described by the receiving efficiency $\eta$. This loss of power results in the SNR of $N_1=10\lg(\eta P_s/P_n)$. Consequently, there is a reduction of SNR, represented by $\Delta N = N_0 - N_1 = -10\lg(\eta)$. The orthogonality of the truncated normalized LG beams $\text{LG}_{m,p_{opt}}(r,\varphi,z_1)\text{circ}(r/R_0(z_1))$, which is registered by the receiver, can be expressed as

$$\left\langle \text{LG}_{m_1,p_{opt}(m_1)}(r,\varphi,z_1)\text{circ}(r/R_0(z_1)) \middle| \text{LG}_{m_2,p_{opt}(m_2)}(r,\varphi,z_1)\text{circ}(r/R_0(z_1)) \right\rangle = \begin{cases} 0 & m_1 \neq m_2 \\ \eta_{max}(m) & m_1 = m_2 = m \end{cases}, \quad (7)$$

where $\text{circ}(\cdot)$ represents the circular function of the receiver aperture. During communication test, the time-varying information carriers in the IRD-OAM multiplexing can be expressed as $\sum_{-m_{max}}^{m_{max}} A_m(t)e^{i\phi_m(t)}\text{LG}_{m,p_{opt}}(r,\varphi,z)$, wherein the received power spectrum of OAM is proportional to $\eta_{max}(m)$ for $m$ in $[-m_{max}, m_{max}]$ with $m_{max}$ denoting the highest OAM value. Hence, the IRD-OAM multiplexing, having the maximum reception efficiency $\eta_{max}(m)$ (or the minimal power loss) for each mode, can minimize the SNR reduction such that $\Delta N(m)$ [$= -10\lg(\eta_{max}(m)$] and thus exceed the information capacity of FSO link systems. For example, the IRD-OAM multiplexing where $m_{max} = 38$, $\eta_{max}(m)$ undergoes a shift from $\eta_{max}(m = 0) = 100\%$ to $\eta_{max}(m = 38) = 10\%$, as shown in Fig. 3**b**. This is due to the limited-size receiver (with a radius of $2.5w(z_1)$) in the FSO link system with $S^2=6.25$. Consequently, $\Delta N(m)$ only changes from 0 to 10 dB, in stark contrast to conventional OAM multiplexing where $\Delta N(m)$ varies from 0 to 60 dB for $m = [0, 38]$.

Despite the theoretical channel crosstalk being negligible (as off-diagonal elements in transmission matrix for IRD-OAM multiplexing are zero), differences in the received power of different OAM modes will result in the SNR fluctuation across different subchannels. This leads to an additional amount of crosstalk during experimental demultiplexing, particularly when $m_{max}$ is large (see Supplementary texts 3 and 4 for explanation). To overcome this issue, we construct the orthonormalized subchannels for the IRD-OAM multiplexing by redistributing the transmitting power spectrum of OAM (which is inversely proportional to $\eta_{max}(m)$) as

$$\left\langle \frac{\text{LG}_{m_1,p_{opt}(m_1)}(r,\varphi,z_1)\text{circ}(r/R_0(z_1))}{\sqrt{\eta_{max}(m_1)}} \middle| \frac{\text{LG}_{m_2,p_{opt}(m_2)}(r,\varphi,z_1)\text{circ}(r/R_0(z_1))}{\sqrt{\eta_{max}(m_2)}} \right\rangle = \delta_{m_1,m_2}. \quad (8)$$

where $\delta_{m_1,m_2}$ represent the Kronecker delta function. The time-varying information carriers of IRD-OAM multiplexing with the uniformly received power spectrum of OAM can be expressed as $\sum_{-m_{max}}^{m_{max}} A_m(t)e^{i\phi_m(t)}\text{LG}_{m,p_{opt}}(r,\varphi,z)/\sqrt{\eta_{max}(m)}$. As such, the SNR of each subchannel is uniform and equal to the total SNR, which is expressed as $N_0 - \Delta N$, where $\Delta N = -10\lg(\eta_{total}(m_{max}))$ with the total reception efficiency $\eta_{total}(m_{max}) = (2m_{max}+1)/\sum_{-m_{max}}^{m_{max}} \eta_{max}^{-1}(m)$ representing the ratio of the total receiving power (proportional to $2m_{max}+1$) to the total transmitting power (proportional to $\sum_{-m_{max}}^{m_{max}} \eta_{max}^{-1}(m)$).

The limited detection capability of the receiver in the practical FSO will constrain the minimal value of

$\eta_{total}(m_{max})$ and maximal value of $m_{max}$, even though the values of $m$ and $p$ in Eq. (1) theoretically have no upper bound. As an illustration, when $m_{max}$ = 38, the SNR reduction of subchannels of IRD-OAM multiplexing is 6.38 dB (as $\Delta N$= -10lg($\eta_{total}(m_{max})$) and $\eta_{total}(38)$ = 23% in Fig. 3**b**). According to our analysis (see Supplementary text 4 for details), if an SNR of ($N_0$-6.38 dB) or a -16 dB level of channel crosstalk is deemed sufficient for the FSO communication, the resulting number of independent subchannels could reach up to $Q_{IRD-OAM}$= 2$m_{max}$+1 = 77. This presents a significant improvement over the conventional OAM multiplexing ($Q_{OAM}$ = 13) and the LG beam multiplexing ($Q_{LG}$ = 21), by achieving 592% and 367%, respectively. These 77 subchannels are shown in Supplementary Movie 1. To further increase the data communication capacity, we can modulate the time-varying amplitude $A_m(t)$ and phase $\phi_m(t)$ using the existing techniques such as quadrature phase-shift keying, amplitude-shift keying and quadrature amplitude modulation. Here we choose to use the binary amplitude-shift keying modulation technology to encode the information into different amplitude sequences of "0"s and "1"s. Besides the existing modulation techniques, the IRD-OAM multiplexing is also compatible with wavelength-division multiplexing and polarization-division multiplexing, resulting in a significant improvement of aggregate capacity limit (e. g., $Q\times2\times T\times$100Gbit/s) of this canonical FSO link system investigated to date[5,11,13-17]. This will be demonstrated by the following FSO experiments.

## Experimental demonstrations of IRD-OAM multiplexing

Figure 4**a** shows the experimental scheme, where there are four key devices including a laser, two spatial light modulators (SLMs) and a camera. The SLM1 (Holoeye GAEA-2, 3.74 μm pixel pitch, 3840×2160) transmits the multiplexed beams based on the information carrier of $\sum_{-m_{\max}}^{m_{\max}} A_m(t)e^{i\phi_m(t)} \mathrm{LG}_{m,p_{opt}(m)}(r,\varphi,z=0)/\sqrt{\eta_{\max}(m)}$ with $w_0$=374 μm, $z_0$=0.83 m, and $\lambda$=532 nm at the waist plane $z$=0. The multiplexed beams propagate in free space for a link distance of 3$z_0$ = 2.49 m and are partially received by the SLM2 (Holoeye Leto, 6.6 μm pixel pitch, 1920 ×1080) due to its limited size; and the recordable beam size depends on the mode order and link distance. For demultiplexing, the SLM2 is imprinted with a truncated Dammann grating pattern[17] produced from the conjugated LG modes, $\sum_{-m_{\max}}^{m_{\max}} \mathrm{LG}^*_{m,p_{opt}(m)}(r,\varphi,z=3z_0)\exp(i2\pi u_m x + i2\pi v_m y)\mathrm{circ}(r/R_0(3z_0))$. The SLM2 diffracts each multiplexed subchannel into a different direction and then is then focused by a lens (focal length of 0.6 m) on the distinct area of a camera (PCO. edge 4.2 bi, 6.5 μm pixel pitch, 2048 ×2048), accomplishing the OAM demultiplexing. A circular mask with a radius of 2.5$w_1$(3$z_0$) = 3 mm acting as the receiver aperture (Fig. 4**a2**) is imprinted on the SLM2, thus simulating a FSO link system of $S^2$ = 6.25. We use the integral intensity over all the pixels of the accumulated region (yellow-dotted circle in Fig. 4**a5**) in the central area of each demultiplexing subchannels to identify whether the bit values are 1 or 0 based on a predetermined discrimination threshold.

We first tested the independent addressability of subchannels for the IRD-OAM multiplexing with $m_{max}$ = 38 (SNR of $N_0$-6.38 dB) in two segments: an 8-bit binary amplitude-shift keying en/decoding IRD-OAM multiplexing for |$m$|∈[31, 38] that assigned bits 1-8 to ($m$, $p_{opt}(m)$) = (-38, 21), (37, 20), (-36, 18), (35, 17), (-34, 16), (33, 15), (-32, 14), (31, 13), respectively, and a 16-bit binary amplitude-shift keying en/decoding IRD-OAM multiplexing for |$m$|∈[0, 30] that assigned bits 1-16 to ($m$, $p_{opt}(m)$) = (-30, 12), (28, 10), (-26, 8), (24, 6), (-22, 5), (20, 4), (-18, 3), (16, 2), (-14, 1), (12, 0), (-10, 0), (8, 0), (-6, 0), (4, 0), (-2, 0), (0, 0), respectively. The experimental results of 256 cases of the 8-bit IRD-OAM multiplexing are shown in Figs. 4**b-d**, exhibiting a bit error rate (BER) of zero via a discrimination threshold in [0.2, 0.4]. The decoding results of the 16-bit IRD-OAM multiplexing are shown in Figs. 4**e-h**.

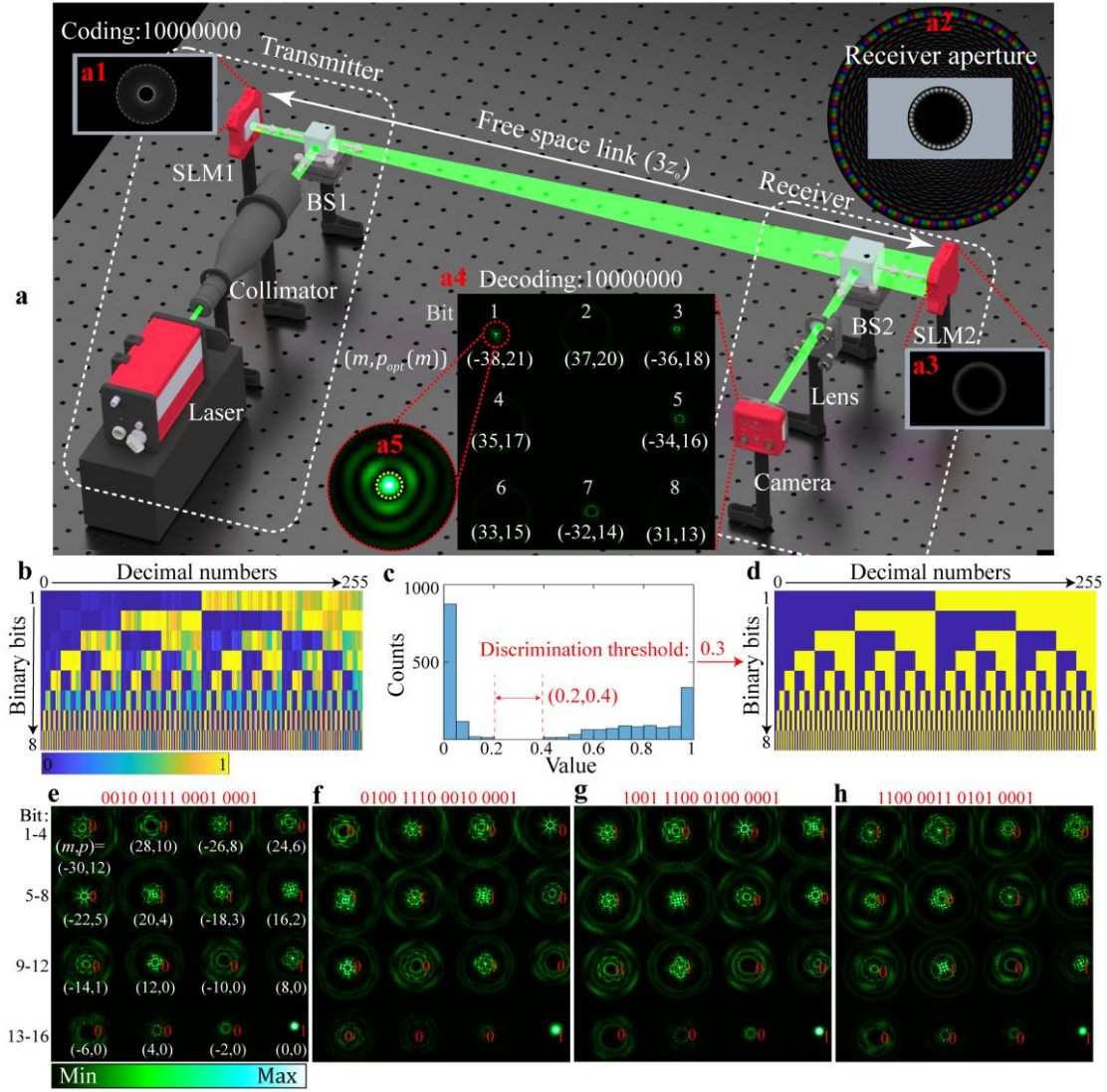

**Fig. 4. Experimental demonstration of IRD-OAM multiplexing. a** Experimental setup: BS1-2, beam splitters; SLM1-2, reflective phase-only spatial light modulators. The insets: **a1**, the complex distribution of LG$_{-38, 21}$($r$, $\varphi$, 0) imprinted on the SLM1; **a2**, the complex distribution of LG$_{-38, 21}$($r$, $\varphi$, 3$z_0$) truncated by the receiver aperture circ($r/R_0$(3$z_0$)) of the SLM2; **a3**, the complex distribution of truncated Dammann grating $\sum_{-m_{max}}^{m_{max}} \text{LG}^*_{m,p_{opt}(m)}(r,\varphi,z=3z_0)\exp(i2\pi u_m x+i2\pi v_m y)\text{circ}(r/R_0(3z_0))/\sqrt{\eta_{max}(m)}$ imprinted on the SLM2. **a4**, the demultiplexing intensity pattern on the camera. **a5**, the zoomed-in view of bit 1 on **a4** with a yellow-dotted circle depicting the accumulated region of the central area. **b** Accumulated values (a.u.) of the accumulated region of 256 cases in the 8-bit IRD-OAM multiplexing shown in **a4**. **c** Corresponding distribution histograms of **b**. **d** Binarized results of **b** with discrimination threshold 0.3. Supplementary Movie S2 shows the experimental demultiplexing intensity patterns of the 256 cases of **b-d**. **e-h** Experimental decoded intensity patterns of 16-bit IRD-OAM multiplexing within |$m$|∈[0, 30] of decimal numbers 10001, 20001, 40001, and 50001, respectively; the white and red font digits indicate the modes and the bit values.

Next, we experimentally demonstrated a practical example of a true color image transmission with 24-bit color depth and $2^{24}$ colors in 512 ×512 pixels (Starry Night by Vincent van Gogh 1889), as shown in Fig. 5**a**. This true color image was divided into three layers: Red, Green, and Blue (RGB) as shown in Fig. 5**b**, with each layer being encoded by an 8 bits color depth (color value: 0-255). Therefore, the information on each pixel of the true color image was transferred through the 24-bit IRD-OAM multiplexing (24 channels are the combination of 8-bit and 16-bit IRD-OAM multiplexing in Fig. 4) for the ultra-high color fidelity. After receiving and decoding the high-density data streams of 512×512×24 bits from the experimental setup of Fig. 4, we recovered the true color image with an ultra-high color fidelity (with a BER = 2.34×10$^{-5}$, much lower than the forward error correction limit[28] of 3.8×10$^{-3}$), as shown in Fig. 5**c**. Figure 5**d** shows the color distribution histograms of this true color image with 24 bits color depth and the RGB layers with 8 bits color depth. The channel crosstalk is below -20 dB (Supplementary text 4) and the experimental SNR [($N_0$-6.38) dB, theoretically] of each channel is above 20 dB with the laser output power of 400 mW (Coherent Verdi G5: 5W max power) and the camera cooled at -20°C (PCO. edge 4.2 bi: 16-bit gray value, 95% QE, 0.2 e-/pixel/s dark current, air and water cooled with the lowest temperature -40°C). By increasing the laser power (i.e., increasing $P_s$) and lowering the detector temperature (i.e., decreasing $P_n$) in our experiment, a higher initial SNR [$N_0 =10\lg(P_s/P_n)$] can be achieved. This, in turn, permits a larger SNR reduction and enables a much larger $m_{max}$ as well as $Q_{IRD-OAM} = 2m_{max}+1$, thereby enhancing the information transfer capacity. This unique ability to exploit SNR resources effectively overcomes the capacity limit of current mode-division multiplexing techniques.

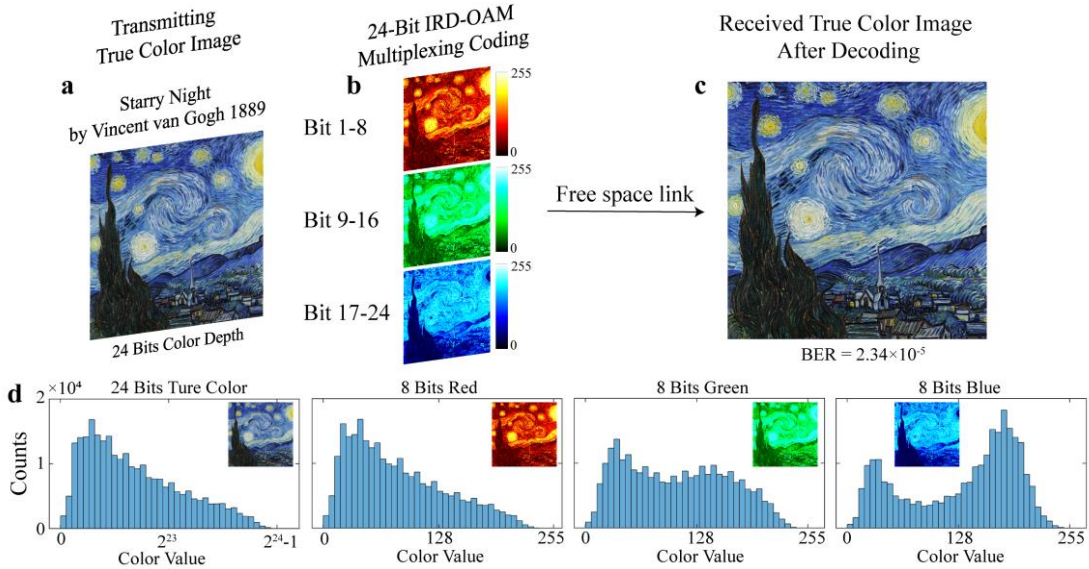

**Fig. 5. Image transmission by 24-bit IRD-OAM multiplexing with ultra-high color fidelity. a** True color image, Starry Night by Vincent van Gogh 1889, with 24 bits color depth and 2$^{24}$ colors including 512 ×512 pixels; **b** Three RGB layers of **a** and were encoded from Bit 1 to 24; **c** Received true color image after recovering with a BER = 2.34×10$^{-5}$. **d** Color distribution histograms of the true color image with 24 bits color depth and the RGB layers with 8 bits color depth.

# Discussion

Diffraction-induced beam size divergence, which increases with mode order and free-space propagation distance, poses a challenge to achieving the high capacity, the miniature receiver, and the long link distance in the FSO systems while operating within physical resource constraints. In practical systems, a trade-off between these three parameters is often necessary, as illustrated by the "triangle of frustration" concept depicted in Fig. 6a. For instance, the base side of the triangle in Fig. 6**a** indicates that reducing the receiver size and increasing the link distance would result in the lower capacity due to the limited physical resources. As shown in Fig. 6, the LG beam multiplexing, by exploiting both the azimuthal and radial degrees of freedom ($r$, $\varphi$), offers a communication ability better than OAM multiplexing, which only utilizes the azimuthal degree of freedom ($\varphi$); however, much better performance, including the improved capacity, the miniaturized receiver, and the longer link distance is achieved by our IRD-OAM multiplexing technique. In the FSO system illustrated in Fig. 4 (with $S^2$ = 6.25, a receiver aperture of $R_0$ = 3 mm, and a link distance of $3z_0$ = 2.49 m), the experimental SNR achieved for the true color image transmission example in Fig. 5 was higher than 20 dB, thereby allowing for a larger reduction, e. g. $\Delta N$=10 dB, which yields $N_0$-$\Delta N$ > 16 dB. Given that $\Delta N$ = -10lg($\eta_{total}(m_{max})$) is equal to 10 dB when $m_{max}$ = 80 [$\eta_{total}(80)$=10% (Supplementary text 5)], we can obtain an ultimate capacity limit of $Q_{IRD-OAM}$ = $2m_{max}$+1 = 161 with channel crosstalk at approximately -13 dB (Supplementary text 4), representing a 1238% improvement over the conventional OAM multiplexing ($Q_{OAM}$=13), as demonstrated in Fig. 6**b**. Moreover, for $\Delta N$ = 10 dB and a desired capacity of $Q$=21, the IRD-OAM multiplexing requires only a minimum quality factor of 0.72 for the FSO system, which is 7.2% of that for the conventional OAM multiplexing with $S^2$ =10 in Fig. 6**c** (see Supplementary text 6 for details). Based on $S^2$=[$R_0$/ $w(z_1)$]$^2$, this significant reduction in $S^2$ enables a longer communication distance $z_1$ and a smaller receiver size $R_0$ for a more miniaturized and integrated FSO device. For instance, for a fixed link distance of $3z_0$ = 2.49 m, the IRD-OAM multiplexing has a minimum receiver aperture of $R_0$ = $2.69w_0$ = 1 mm, which is 26.9% of that required for the OAM multiplexing with $R_0$ = $10w_0$ = 3.7 mm in Fig. 6**d**; in contrast, for a fixed receiver aperture of $R_0$ = $8w_0$ = 3 mm, the IRD-OAM multiplexing has a maximum link distance of $z_1$ = $9.38z_0$ = 7.79 m, which is 403% of that required for the OAM multiplexing with $z_1$ = $2.33z_0$ = 1.93 m in Fig. 6**e**. Furthermore, increasing the laser power (e. g. from 400 mW to 5 W) and lowering the detector temperature (e. g. from -20°C to -40°C) in our experiment can result in a much higher SNR (as the SNR $N_0$ =10lg($P_s$/$P_n$)), allowing a larger SNR reduction of $\Delta N$ and resulting in a higher $Q_{IRD-OAM}$ = $2m_{max}$+1. This, in turn, would significantly enhance the capacity, reduce the receiver, and extend the link distance further.

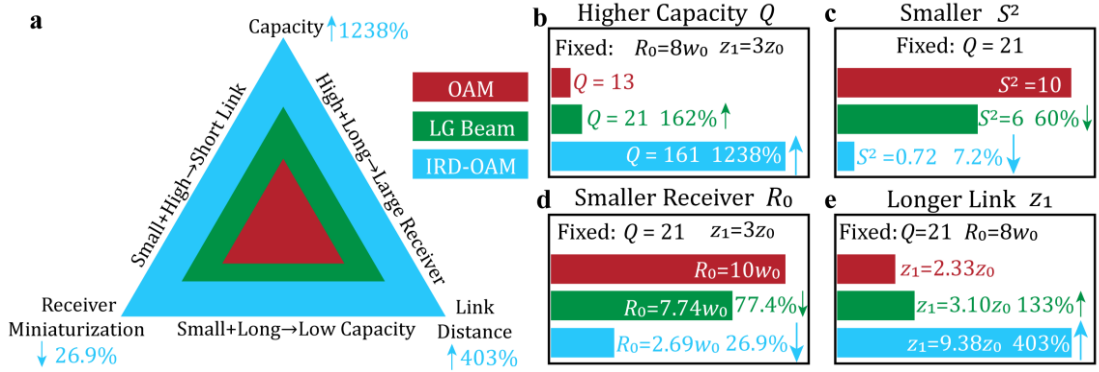

**Fig. 6. Ultimate limits of capacity, distance, and receiver miniaturization. a** The "triangle of frustration" among capacity, receiver miniaturization and link distance of the FSO system; the larger the triangle area, the better the communication performance with the FSO system. The comparison among the OAM multiplexing (the red bars), the LG beam multiplexing (the green bars), and the IRD-OAM multiplexing with SNR = ($N_0$-10) dB (the blue bars). **b** The numbers of independently addressable subchannels $Q$ with the fixed receiver aperture of $R_0$=$8w_0$ and link distance $z_1$ = $3z_0$. **c** The smallest system quality factors $S^2$ for a desired number of subchannels of $Q$ = 21. **d** The smallest receiver apertures for a required number of subchannels of $Q$ = 21 with the fixed link distance of $z_1$ = $3z_0$. **e** The longest link distances for a required number of subchannels of $Q$ = 21 with the fixed receiver aperture of $R_0$=$8w_0$.

In the conventional mode-division multiplexing techniques such as the OAM multiplexing and the LG beam multiplexing, increasing the laser power or detector sensitivity enhances the SNR but cannot overcome the limits in capacity, link distance and receiver miniaturization. In contrast, the IRD-OAM multiplexing enables us to surpass these initial limits of the FSO system by fully exploiting SNR resources without requiring additional hardware modifications. Specifically, increasing the laser power and detector sensitivity for a higher SNR [$\uparrow N_0 = 10\lg(\uparrow P_s / \downarrow P_n)$] in the IRD-OAM multiplexing allows a larger reduction of $\Delta N$, i.e. increasing the detection redundancy can enhance the capacity, miniaturize the receiver and extend the link distance. Hence, the performance of the IRD-OAM multiplexing is much superior to the OAM multiplexing and the LG beam multiplexing in the same FSO system, as illustrated by the blue triangle with the largest area shown in Fig. 6**a**.

## Conclusion

In summary, the LG beams suffer from divergence due to diffraction during free-space propagation, limiting the link distance and capacity of the FSO communication with a given receiver size[20,21]. Instead of using the entire beam profile, the IRD-OAM multiplexing, however, recovers the information from the innermost ring with a limited-size receiver, providing superior dynamic transmission characteristics such as a much smaller quality factor, self-healing capability with sidelobe assistance, and better identification ability (owing to higher intensity and phase gradient of the innermost-ring than the sidelobes). Unlike the OAM multiplexing and the LG beam multiplexing, which only trade off among capacity, link distance, and receiver miniaturization, the IRD-OAM multiplexing can significantly expand the "triangle of frustration" to greatly exceed the original information transfer limits of the FSO systems without additional hardware modifications. The proposed IRD-OAM multiplexing technique can fully integrate and collaborate on multidimensional physical resources of the FSO systems, making it more adaptable for designing the future FSO communication systems with ultra-high capacity, ultra-long distance, and highly-integrated devices for deep-space, near-Earth and Earth-surface applications.

## Reference

1   Shannon, C. E. A mathematical theory of communication. *ACM SIGMOBILE mobile computing and communications review* **5**, 3-55 (2001).

2   Hanzo, L., Ng, S. X., Webb, W. & Keller, T. *Quadrature amplitude modulation: From basics to adaptive trellis-coded, turbo-equalised and space-time coded OFDM, CDMA and MC-CDMA systems*.   (IEEE Press-John Wiley, 2004).

3   Evangelides, S. G., Mollenauer, L. F., Gordon, J. P. & Bergano, N. S. Polarization multiplexing with solitons. *J. Lightwave Technol.* **10**, 28-35 (1992).

4   Mukherjee, B. *Optical WDM networks*.   (Springer Science & Business Media, 2006).

5   Wang, J. *et al.* Terabit free-space data transmission employing orbital angular momentum multiplexing. *Nat. Photonics* **6**, 488-496 (2012).

6   Bozinovic, N. *et al.* Terabit-scale orbital angular momentum mode division multiplexing in fibers. *Science* **340**, 1545-1548 (2013).

7   Wan, Z. *et al.* Divergence-degenerate spatial multiplexing towards future ultrahigh capacity, low error-rate optical communications. *Light: Science & Applications* **11**, 1-11 (2022).

8   Wang, J. *et al.* in *2014 The European Conference on Optical Communication (ECOC).*   1-3 (IEEE).

9   Wang, J. *et al.* Orbital angular momentum and beyond in free-space optical communications. *Nanophotonics* **11**, 645-680 (2022).

10  Willner, A. E. *et al.* Optical communications using orbital angular momentum beams. *Advances in optics and photonics* **7**, 66-106 (2015).

11  Huang, H. *et al.* 100 Tbit/s free-space data link enabled by three-dimensional multiplexing of orbital angular momentum, polarization, and wavelength. *Opt. Lett.* **39**, 197-200 (2014).
12  Djordjevic, I. B. Deep-space and near-Earth optical communications by coded orbital angular momentum (OAM) modulation. *Opt. Express* **19**, 14277-14289 (2011).
13  Gibson, G. *et al.* Free-space information transfer using light beams carrying orbital angular momentum. *Opt. Express* **12**, 5448-5456 (2004).
14  Wang, J. *et al.* in *IEEE Photonic Society 24th Annual Meeting.*    587-588 (IEEE).
15  Wang, J. *et al.* in *European Conference and Exposition on Optical Communications.*    We. 10. P11. 76 (Optical Society of America).
16  Fazal, I. M. *et al.* in *Frontiers in Optics.*    FTuT1 (Optica Publishing Group).
17  Lei, T. *et al.* Massive individual orbital angular momentum channels for multiplexing enabled by Dammann gratings. *Light: Science & Applications* **4**, e257-e257 (2015).
18  Willner, A. E. *et al.* Design challenges and guidelines for free-space optical communication links using orbital-angular-momentum multiplexing of multiple beams. *Journal of Optics* **18**, 074014 (2016).
19  Xie, G. *et al.* Performance metrics and design considerations for a free-space optical orbital-angular-momentum–multiplexed communication link. *Optica* **2**, 357-365 (2015).
20  Krenn, M. *et al.* Communication with spatially modulated light through turbulent air across Vienna. *New Journal of Physics* **16**, 113028 (2014).
21  Krenn, M. *et al.* Twisted light transmission over 143 km. *Proceedings of the National Academy of Sciences* **113**, 13648-13653 (2016).
22  Zhao, N., Li, X., Li, G. & Kahn, J. M. Capacity limits of spatially multiplexed free-space communication. *Nat. Photonics* **9**, 822-826 (2015).
23  Phillips, R. L. & Andrews, L. C. Spot size and divergence for Laguerre Gaussian beams of any order. *Appl. Opt.* **22**, 643-644 (1983).
24  Siegman, A. E. in *Optical resonators.*    2-14 (Spie).
25  Padgett, M. On the focussing of light, as limited by the uncertainty principle. *J. Mod. Opt.* **55**, 3083-3089 (2008).
26  Vaity, P. & Rusch, L. Perfect vortex beam: Fourier transformation of a Bessel beam. *Opt. Lett.* **40**, 597-600 (2015).
27  Mendoza-Hernández, J., Hidalgo-Aguirre, M., Ladino, A. I. & Lopez-Mago, D. Perfect Laguerre–Gauss beams. *Opt. Lett.* **45**, 5197-5200 (2020).
28  Richter, T. *et al.* Transmission of single-channel 16-QAM data signals at terabaud symbol rates. *J. Lightwave Technol.* **30**, 504-511 (2011).

## Data availability

All data that support the findings of this study are available within the article and Supplementary Information, or available from the corresponding author on reasonable request.

## Acknowledgements

This work was financially supported by National Key Research and Development Program of China (2022YFA1404800, 2018YFA0306200) and National Natural Science Foundation of China (12234009, 12274215, 11922406).

## Author contributions

W.Y. and J. D. proposed the original idea and designed the study. W. Y. built the experimental system and performed the experiments. Y.G., X.L., and Z.Y. assisted the experiments. J.D. and H.T.W. supervised the project. All authors contributed to writing the manuscript.

## Competing interests

The authors declare no competing interests